\documentclass[preprint,showpacs,preprintnumbers,amsmath,amssymb,prb]{revtex4-1}

\usepackage[dvipdfmx]{graphicx}
\usepackage{dcolumn}
\usepackage{bm}
\usepackage{mhchem} 
\usepackage{ascmac}
\usepackage{color} 
\usepackage[normalem]{ulem}
\begin{document}
%
\title{Method of images for one-dimensional discrete random walk under a reflecting barrier}
\author{
Kazuhiko Seki
}
\email{k-seki@aist.go.jp}
\affiliation{GZR, National Institute of Advanced Industrial Science and Technology (AIST), Onogawa 16-1 AIST West, Ibaraki 305-8569, Japan
}
\date{\today}
\begin{abstract}

The transition probability for a one-dimensional discrete symmetric random walk under a reflecting barrier was once given by the method of images.  [S. Chandrasekhar, Rev. Mod. Phys. 15, 1 (1943).]
However, several inconsistencies have been reported when the method of images is applied in cases where a reflecting barrier is considered, even after the exact solution has been obtained. 
Here, we explicitly show that the method of images becomes applicable if the image position is shifted. 
We also discuss the implication to the fluctuation theorem in infinite discrete lattice models. 
\end{abstract}

\maketitle

\section{Introduction} 
\label{sec:intro}
In a well-known review article, Chandrasekhar presented the transition probability (Green function) constructed by the method of images for a reflecting boundary condition as well as for an absorbing boundary condition. \cite{Chandrasekhar_43}
Chandrasekhar's transition probability for a reflecting boundary condition led to confusion;  
many authors, including Goodrich, \cite{Goodrich_54} accepted it, whereas others noted several inconsistencies in it. \cite{RAZINAQVI_82,Orlowski_03,Hansen_11} 
An attempt to modify the boundary condition to match Chandrasekhar's transition probability has been reported. \cite{Hansen_11} However, even when this modification is adopted, the transition probability at the reflecting site with the modified transition rate constant should be set to be one-half of Chandrasekhar's transition probability. \cite{Hansen_11} van Kampen and Oppenheim derived the exact solution under a reflecting 
boundary condition using normal modes; however, the solution involves an infinite summation. \cite{vanKampen_72}
van Kampen and Oppenheim also introduced an approximation applicable when the lattice spacing is smaller than the distances of interest and noted that the probability density no longer has a zero slope at the site where the 
reflecting boundary condition is imposed. \cite{vanKampen_72} 
Researchers have also obtained the exact solution by solving a recursive equation and have given an intuitive interpretation based on the method of images. \cite{Montroll_67} However, the shift of the image was not mathematically elucidated. 
The method of images is attributed to Lord Kelvin in Feller's book.~\cite{Feller_70}  
Feller applied the method to present solutions to the symmetric random walk problem that correspond to both reflecting and absorbing boundary conditions.  
Notably, the solution corresponding to the reflecting boundary condition involves a shift in the image position, which contradicts Chandrasekhar's formulation.  
However, no explanation for this shift is provided in the book.

Here, 
we employ the method of images, initially allowing the image position to remain arbitrary. 
The image position is then determined by imposing the reflecting boundary condition, 
which necessitates a shift from the naively assumed location. 
This approach offers a simpler alternative to directly solving the continuous-time discrete lattice random walk with a reflecting boundary. 
Furthermore, we show that Chandrasekhar’s original expression can be modified to accommodate the reflecting boundary condition. 
The transition probability obtained using the shifted image satisfies the fluctuation theorem, whereas that derived from Chandrasekhar’s formulation violates it. 
In the continuum limit, both formulations converge to a unique solution consistent with the fluctuation theorem. 
We further discuss the implications of the fluctuation theorem for infinite discrete lattice systems, including those exhibiting asymmetry (biased random walks).

\section{Boundary free random walk} 
\label{sec:free}

As a preliminary step, we first present a known result for an unbiased random walk in an infinite one-dimensional lattice without a boundary. 
The result helps to construct the transition probability using the method of images by properly locating the image position to satisfy the reflecting boundary condition. 
We consider a random walk along the $x$-axis, with the lattice spacing denoted by $b$. 
When the total transition rate constant is denoted by $\gamma$, the transition probability for a particle at $x$ at time $t \sim t+dt$ starting from the initial position $x_{\rm i}$ [$G_0(x,x_{\rm i}, t)$] in an unbiased random walk in an infinite one-dimensional lattice satisfies
\begin{align}
\frac{\partial}{\partial t} G_0(x,x_{\rm i}, t)=\frac{\gamma}{2} G_0(x-b,x_{\rm i}, t)-\gamma G_0(x,x_{\rm i}, t)+\frac{\gamma}{2} G_0(x+b,x_{\rm i}, t) .
\label{eq:unbiased}
\end{align}
We denote the Laplace transform of the time-dependent function $f(t)$ by $\hat{f}(s)$. 
The initial condition is given by  $\delta_{x, x_{\rm i}}$, where 
$\delta_{x,y}$ denotes the Kronecker delta, which equals 1 if $x = y$ and 0 otherwise. 
As in Appendix A, the solution of Eq. (\ref{eq:unbiased}) with the initial condition $\delta_{x, x_{\rm i}}$ for one-dimensional unbounded space
[$G_0(x,x_{\rm i},t)$] in the Laplace domain can be expressed as 
\begin{align}
\hat{G}_0(x,x_{\rm i},s)&= \frac{b}{2\pi} \int_{-\pi/b}^{\pi/b} dk\, \frac{\exp[-ik(x-x_{\rm i})]}{s+\gamma\left[1-\cos\left(kb \right)\right]}. 
\label{eq:G0_0d}
\end{align}
By noting that $G_0(x,x_{\rm i}, t)$ is a function of $x-x_{\rm i}$ alone, we express it as $G_0(x-x_{\rm i}, t)\equiv G_0(x,x_{\rm i}, t)$ when needed. 
By the inverse Laplace transform of Eq. (\ref{eq:G0_0d}), $G_0(x-x_{\rm i}, t)$ is obtained as
\begin{align}
G_0(x-x_{\rm i}, t)&= \frac{b}{2\pi} \int_{-\pi/b}^{\pi/b} dk\, \cos[k(x-x_{\rm i})]
\exp\left[- \gamma\left(1-\cos\left(kb \right)\right)t
\right]
\label{eq:G0_0t_v1}\\
&=\exp\left(-\gamma t \right) I_{|x-x_{\rm i}|/b}\left(\gamma t \right), 
\label{eq:G0_0t_v2}
\end{align}
where the integral representation of the modified Bessel function of the first kind [$I_n (z)$] for an integer index is used. 
Equation (\ref{eq:G0_0t_v2}) is known expression given by Eq. (154) of Ref. \citenum{Montroll_67} 
if symmetric random walk is considered. 

\section{Method of images} 
\label{sec:fMI}

\begin{figure}[h]
\begin{center}
\includegraphics[width=10cm]{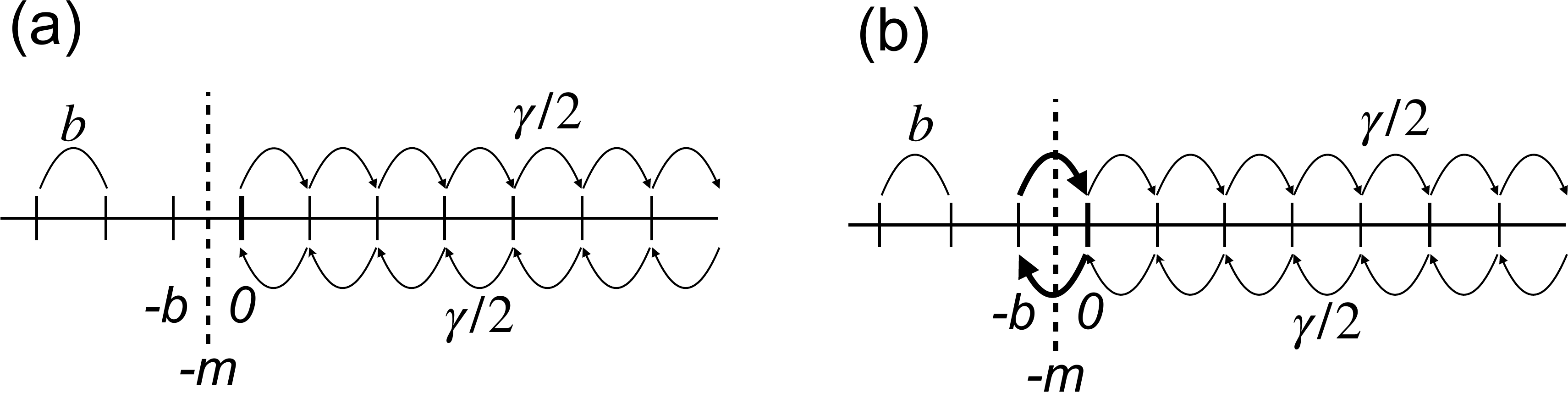}
\end{center}
\caption{Schematics of random walk model with reflecting boundary located at $-m$. The lattice spacing is denoted by $b$. 
The total transition rate is denoted by $\gamma$. 
(a) The reflecting boundary is located at $x=0$. 
(b) The boundary condition $G_{\rm rfl}(-b,x_{\rm i}, t)=G_{\rm rfl}(0,x_{\rm i}, t)$ is imposed at the auxiliary site $-b$. 
The boundary condition at the auxiliary site [(b)] is equivalent to the reflecting 
boundary condition [(a)]. \cite{vanKampen_72,vanKampen}
}
\label{fig:1}
\end{figure}

Now, we construct the transition probability using the method of images 
[$G_{\rm rfl}(x,x_{\rm i}, t)$] to satisfy the reflecting boundary condition at site $0$:  \cite{vanKampen_72,vanKampen}
\begin{align}
\frac{\partial}{\partial t} G_{\rm rfl}(0,x_{\rm i}, t)=-\frac{\gamma}{2} G_{\rm rfl}(0,x_{\rm i}, t)+\frac{\gamma}{2} G_{\rm rfl}(b,x_{\rm i}, t) .
\label{eq:ubnaturalbc}
\end{align} 
As shown in Fig. \ref{fig:1} (a), the reflecting boundary condition in Eq.~(\ref{eq:ubnaturalbc}) ensures that probability flux at the boundary vanishes, meaning a random walker cannot cross the origin ($x=0$) but is instead reflected. 
The initial condition is given by $\delta_{x, x_{\rm i}}$ with $x_{\rm i}\geq 0$ and we consider a random walk over $x=,0,b,2b,\cdots$. 
By considering a random walk over $x=-b,0,b,2b,\cdots$, the reflecting boundary condition is known to be equivalent to considering an auxiliary boundary condition at site $-b$: \cite{vanKampen_72,vanKampen}
\begin{align}
\frac{\gamma}{2}  G_{\rm rfl}(-b,x_{\rm i}, t)=
\frac{\gamma}{2} G_{\rm rfl}(0,x_{\rm i}, t). 
\label{eq:ubaux}
\end{align}
By adding Eq. (\ref{eq:ubaux}) with 
\begin{align}
\frac{\partial}{\partial t} G_{\rm rfl}(0,x_{\rm i}, t)=\frac{\gamma}{2} G_{\rm rfl}(-b,x_{\rm i}, t)-\gamma G_{\rm rfl}(0,x_{\rm i}, t)+\frac{\gamma}{2} G_{\rm rfl}(b,x_{\rm i}, t) , 
\label{eq:unbiased0}
\end{align}
we obtain the 
reflecting boundary condition given by Eq. (\ref{eq:ubnaturalbc}).  
Therefore, imposing the 
reflecting boundary condition given by Eq. (\ref{eq:ubnaturalbc}) is equivalent to imposing Eq. (\ref{eq:ubaux}) at the auxiliary site $-b$ (see Fig. \ref{fig:1}). \cite{vanKampen_72,vanKampen}
This auxiliary formulation simplifies the analysis by circumventing the need to explicitly modify the time evolution equation at the boundary.

Instead of directly considering the 
reflecting boundary condition, we consider the auxiliary boundary condition given by Eq. (\ref{eq:ubaux}): 
$G_{\rm rfl}(-b,x_{\rm i}, t)=G_{\rm rfl}(0,x_{\rm i}, t)$.  
We study exact solutions of Eq. (\ref{eq:unbiased}) for the initial condition,  $\delta_{x, x_{\rm i}}$ with $x_{\rm i}\geq 0$, under the reflecting boundary condition [Eq. (\ref{eq:ubnaturalbc})] 
by considering a random walk over $x=-b,0,b,2b,\cdots$ under the boundary condition at site $-b$ [Eq. (\ref{eq:ubaux})]. 

Alternative forms of the reflecting boundary condition have been proposed in Refs.~\citenum{vanKampen} and \citenum{Kac_47}, in which the time evolution equations at and near the boundary are modified to ensure consistency with reflection:
\begin{align}
\frac{\partial}{\partial t} G_{\rm rfl}(1,x_{\rm i}, t) &= \gamma G_{\rm rfl}(0,x_{\rm i}, t) - \gamma G_{\rm rfl}(b,x_{\rm i}, t) + \frac{\gamma}{2} G_{\rm rfl}(2b,x_{\rm i}, t), 
\label{eq:refleBCalter1}\\
\frac{\partial}{\partial t} G_{\rm rfl}(0,x_{\rm i}, t) &= -\gamma G_{\rm rfl}(0,x_{\rm i}, t) + \frac{\gamma}{2} G_{\rm rfl}(b,x_{\rm i}, t).
\label{eq:refleBCalter2}
\end{align}
We adopt Eq.~(\ref{eq:ubnaturalbc}), known as the "pure reflecting boundary condition" (Sec.~VI.7 of Ref.~\citenum{vanKampen}), instead of the more complex "impure reflecting boundary condition" given by Eqs.~(\ref{eq:refleBCalter1})--(\ref{eq:refleBCalter2}).
By "exact solution", we refer to solving Eq.~(\ref{eq:unbiased}) with $\delta_{x, x_{\rm i}}$ ($x_{\rm i} \geq 0$) under the pure reflecting boundary condition [Eq.~(\ref{eq:ubnaturalbc})].

By considering the reflecting boundary located at site $0$ and the auxiliary boundary located at site $-b$, the location of the reflecting boundary might be between site $0$ and site $-b$. 
We denote the location of the reflecting boundary by $-m$ [see Fig. \ref{fig:1} (b)].
By noting that \(-m - [x_{\rm i} - (-m)] = -x_{\rm i} - 2m\), the transition probability constructed using the method of images can be written as
\begin{align}
G_{\rm rfl}(x, x_{\rm i}, t) \equiv G_0(x, x_{\rm i}, t) + G_0(x, -x_{\rm i} - 2m, t),
\label{eq:Greenrfll}
\end{align}
where \(m\) is chosen to satisfy the pure reflecting boundary condition given by Eq.~(\ref{eq:ubnaturalbc}). 
With this choice, \(G_{\rm rfl}(x, x_{\rm i}, t)\) represents the exact solution under the reflecting boundary condition.
In the boundary condition given by Eq. (\ref{eq:ubaux}), we have 
$G_{\rm rfl}(-b,x_{\rm i}, t)=G_0(x_{\rm i}+b, t)+G_0(x_{\rm i}-b+2m, t)$ and
$G_{\rm rfl}(0,x_{\rm i}, t)=G_0(x_{\rm i}, t)+G_0(x_{\rm i}+2m, t)$. 
Using Eq. (\ref{eq:G0_0d}) and $\cos(A)+\cos(B)=2 \cos[(A+B)/2]\cos[(A-B)/2]$, we obtain
\begin{align}
\hat{G}_{\rm rfl}(-b,x_{\rm i},s)&= \frac{b}{2\pi} \int_{-\pi/b}^{\pi/b} dk\, \frac{2\cos[k(x_{\rm i} + m)]\cos[k(b-m)]}{s+\gamma\left[1-\cos\left(kb \right)\right]}, 
\label{eq:unbiasedrfllmb}\\
\hat{G}_{\rm rfl}(0,x_{\rm i},s)&= \frac{b}{2\pi} \int_{-\pi/b}^{\pi/b} dk\, \frac{2\cos[k(x_{\rm i} + m)]\cos(km)}{s+\gamma\left[1-\cos\left(kb \right)\right]}.  
\label{eq:unbiasedrfll0}
\end{align}
We note that Eq. (\ref{eq:unbiasedrfllmb}) is equivalent to Eq. (\ref{eq:unbiasedrfll0}) when $m=b/2$ is set; the boundary condition given by Eq. (\ref{eq:ubaux}) is therefore satisfied. 
As a result, 
\begin{align}
G_{\rm rfl}(x,x_{\rm i},t)\equiv G_0(x,x_{\rm i}, t)+G_0(x,-x_{\rm i}-b, t)
\label{eq:Grflsol}
\end{align}
 is the solution of the unbiased lattice random walk with the reflecting boundary condition at site $0$. 
As in Appendix A, the equivalent result can be obtained for discrete-time random walks (DTRW). 
In contrast, Chandrasekhar's solution corresponds to the case \(m = 0\) in Eq.~(\ref{eq:Greenrfll}) and therefore does not satisfy the pure reflecting boundary condition defined by Eq.~(\ref{eq:ubnaturalbc}). 
Using Eq. (\ref{eq:G0_0t_v2}), we can explicitly express $G_{\rm rfl}(x,x_{\rm i},t)$ as
\begin{align}
G_{\rm rfl}(x,x_{\rm i},t)&=\exp\left(-\gamma t \right) \left[I_{|x-x_{\rm i}|/b}\left(\gamma t \right)
+ I_{(x+x_{\rm i}+b)/b}\left(\gamma t \right)
\right] .
\label{eq:Grflsolv1}
\end{align}
For analytical simplicity, we have considered a continuous-time simple random walk; however, Eq. (14) also applies to 
$\hat{G}_0(x,x_{\rm i},s)$ as defined in Eq. (\ref{eq:G0_0dAg}) of Appendix A, which corresponds to a continuous-time random walk with a general waiting time distribution, replacing the exponential case given in Eq. (\ref{eq:G0_0d}). \cite{Montroll,Hughes_95}
Previously, van Kampen and Oppenheim showed that the probability density no longer has a zero slope at $x =-b$ (nor at $x=0$), but instead at $x=-b/2$. \cite{vanKampen_72}
Their conclusion is consistent with Eq. (\ref{eq:Grflsol}).  
They also demonstrated that the slip coefficient can be defined at $x = -b$ 
when total probability is not conserved due to absorption at the boundary. 
Interestingly, they argued that the concept of a slip coefficient does not apply in the case of a reflecting boundary, 
since the probability density exhibits a zero slope at $x = -b/2$ rather than at $x = -b$; 
in other words, the effective reflecting boundary is shifted from the actual boundary location.
During the final drafting of this article, we became aware that Eq. (\ref{eq:Grflsol}) was proved recently in a very different manner in a different context. \cite{Corwin_18}
Previously, Eq. (\ref{eq:Grflsolv1}) was derived by solving a recurrence relation, and its connection to the method of images was noted with the remark, ``This could have been derived directly by the theory of images.'' \cite{Montroll_67}
The boundary condition given by $G_{\rm rfl}(-b,x_{\rm i}, t)=G_{\rm rfl}(0,x_{\rm i}, t)$ implies that a walker just arriving at the auxiliary site is placed back at the physical site located at $0$. It also implies that the rational position of the reflecting barrier, although we are considering the discrete lattice, might be the midpoint between the auxiliary site at $-1$ and the physical site at $0$. \cite{Schwarz_75}
Here, we derived mathematically that the reflecting barrier should be located at the site $-1/2$ to satisfy $G_{\rm rfl}(-b,x_{\rm i}, t)=G_{\rm rfl}(0,x_{\rm i}, t)$. 

\section{Discussion and Conclusion} 
\label{sec:DC}

What is the boundary condition satisfied by Chandrasekhar's transition probability?
By setting $m=0$, which is appropriate for Chandrasekhar's transition probability, we find $\hat{G}_{\rm rfl}(-b,x_{\rm i},s)=k_{\rm a}(s) \hat{G}_{\rm rfl}(0,x_{\rm i},s)$  in the Laplace domain with $k_{\rm a}(s) =\hat{G}_{\rm rfl}(-b,x_{\rm i},s)/\hat{G}_{\rm rfl}(0,x_{\rm i},s)$; 
the value of $k_{\rm a}(s)$ satisfies $0<k_{\rm a}(s)<1$ because  of $\cos(kb)<1$ in $\hat{G}_{\rm rfl}(-b,x_{\rm i},s)$ with $m=0$ 
and indicates a partially absorbing boundary condition in the Laplace domain. 
Therefore, Chandrasekhar's transition probability is a solution for the partially absorbing boundary condition  in the Laplace domain. 
The solution for the boundary condition given by $\hat{G}_{\rm rfl}(-b,x_{\rm i},s)=\alpha(s) \hat{G}_{\rm rfl}(0,x_{\rm i},s)$ can be constructed by an infinite summation using $\hat{G}_0(x+x_{\rm i}+b j, s)$ with $j$ being a non-negative integer for $\alpha(s)<1$. \cite{Montroll_67}
Because the equations are linear, the transition probability for a perfectly reflecting boundary condition can be expressed using a linear combination of Chandrasekhar's transition probability and the solution for  
$\alpha(s)=1-k_{\rm a}(s)$. 
[See Appendix B for the correction of Chandrasekhar's expression.] 
The transition probability given by Eq. (\ref{eq:Grflsol}) is invariant under the exchange of $x$ and $x_{\rm i}$ and, therefore, satisfies the fluctuation theorem.  
However,  the transition probability 
constructed on the basis of Chandrasekhar's transition probability violates the fluctuation theorem as shown in Appendix B; \cite{Seki_23} 
the fluctuation theorem is broken owing to the non-uniqueness of the transition probability in infinite discrete lattice models, 
where  the reflecting boundary condition is imposed at one end. 
As demonstrated in Appendix C, the same conclusion applies to DTRWs.
When the continuous limit in space is taken, the solution should converge to a unique solution given by the continuous limit of Eq. (\ref{eq:Grflsolv1}), 
which satisfies the fluctuation theorem. [See Appendix B]

It is straightforward to extend the results presented above to an asymmetric random walk model, as demonstrated in 
Appendix D. 
In the infinite discrete lattice model, certain expressions for the transition probability satisfy the fluctuation theorem, 
while others do not. When the continuous limit in space is applied, a unique solution that satisfies the fluctuation theorem is obtained.

Chandrasekhar also presented the transition probability for a perfectly absorbing boundary condition using the method of images, and the solution should be exact;  indeed, $G_0(\left| x-x_{\rm i} \right|, t)-G_0(x+x_{\rm i}, t)$ should be zero at  $x=0$ (the absorbing site). \cite{Chandrasekhar_43,Montroll_67} 
Note that only the absorbing site is considered when constructing the transition probability to satisfy the perfectly absorbing boundary condition by the method of images. 
However, the transition probability for the perfectly reflecting boundary condition is constructed using the method of images on the basis of the slope of the density at the reflecting site, where the definition of the slope requires two sites. 
This might be the reason for the shift in the transition probability constructed by the method of images when considering the reflecting boundary condition. 

\acknowledgments 
I would like to thank Professor Hisao Hayakawa (Kyoto University) and Professor Yoshihiko Hasegawa (The University of Tokyo) for their interest and for discussion during the workshop on "Frontiers in Non-equilibrium Physics 2024" 
(YITP-T-24-01) at Yukawa Institute for Theoretical Physics (Kyoto University, Japan). I would like to acknowledge the warm hospitality during our stay in the YITP.


\newpage
\renewcommand{\theequation}{A.\arabic{equation}} 
\setcounter{equation}{0}  
\section*{Appendix A. Derivation of Eq. (\ref{eq:G0_0d})}

In the Laplace domain, Eq.~(\ref{eq:unbiased}) can be written as
\begin{align}
s \hat{G}_0(x,x_{\rm i}, s) - \delta_{x, x_{\rm i}} = \frac{\gamma}{2} \hat{G}_0(x-b,x_{\rm i}, s) - \gamma \hat{G}_0(x,x_{\rm i}, s) + \frac{\gamma}{2} \hat{G}_0(x+b,x_{\rm i}, s),
\label{eq:unbiased_Lap}
\end{align}
where $\delta_{x, x_{\rm i}}$ denotes the Kronecker delta, which equals 1 if $x = x_{\rm i}$ and 0 otherwise.

We introduce the discrete Fourier transform \cite{Hughes_95} as
\begin{align}
\tilde{G}_0(k,x_{\rm i}, s) = \sum_x e^{ikx} \hat{G}_0(x,x_{\rm i}, s),
\label{eq:dF}
\end{align}
with its inverse given by \cite{Hughes_95}
\begin{align}
\hat{G}_0(x,x_{\rm i}, s) = \frac{b}{2\pi} \int_{-\pi/b}^{\pi/b} dk\, e^{-ikx} \tilde{G}_0(k,x_{\rm i}, s).
\label{eq:idF}
\end{align}

Applying $\sum_x e^{ikx}$ to both sides of Eq.~(\ref{eq:unbiased_Lap}), we obtain
\begin{align}
s \tilde{G}_0(k,x_{\rm i}, s) - e^{ikx_{\rm i}} = \gamma \left[\cos(kb) - 1\right] \tilde{G}_0(k,x_{\rm i}, s),
\label{eq:unbiased_Lap1}
\end{align}
where we used the identity $\sum_x e^{ikx} \hat{G}_0(x \mp b,x_{\rm i}, s) = \sum_x e^{ik(x \pm b)} \hat{G}_0(x,x_{\rm i}, s)$.

Rearranging Eq.~(\ref{eq:unbiased_Lap1}) and using Eq.~(\ref{eq:idF}), we obtain Eq.~(\ref{eq:G0_0d}) in the form:
\begin{align}
\hat{G}_0(x,x_{\rm i},s) = \frac{b}{2\pi} \int_{-\pi/b}^{\pi/b} dk\, \frac{e^{-ik(x - x_{\rm i})}}{s + \gamma[1 - \cos(kb)]}.
\label{eq:G0_0dA}
\end{align}

Equation~(\ref{eq:G0_0dA}) can also be derived from the known expression for a continuous-time random walk, where the waiting time distribution is denoted by $\psi(t)$. In the Laplace domain, this is expressed as \cite{Hughes_95,Weiss_94}
\begin{align}
\hat{G}_0(x,x_{\rm i},s) = \frac{b}{2\pi} \int_{-\pi/b}^{\pi/b} dk\, \hat{\varphi}(s) \frac{e^{-ik(x - x_{\rm i})}}{1 - \hat{\psi}(s)\cos(kb)},
\label{eq:G0_0dAg}
\end{align}
where $\hat{\psi}(s)$ is the Laplace transform of the waiting time distribution, and $\hat{\varphi}(s)$ is the Laplace transform of the survival probability, given by $\varphi(t) = \int_t^\infty dt'\, \psi(t')$.

Using the identity $\hat{\varphi}(s) = [1 - \hat{\psi}(s)]/s$, we recover Eq.~(\ref{eq:G0_0dA}) for the case of an exponential waiting time distribution, $\psi(t) = \gamma e^{-\gamma t}$, with Laplace transform $\hat{\psi}(s) = \gamma / (s + \gamma)$. 
\cite{Weiss_94}

Before concluding this Appendix, we demonstrate the connection between Eq.~(\ref{eq:unbiased}) and the discrete-time formulation presented by Chandrasekhar~\cite{Chandrasekhar_43}. 
In Chandrasekhar's discrete model, the probability $W(x, x_{\rm i}, n)$ denotes the likelihood that a particle starting at position 
$x_{\rm i}$ reaches position $x$ after $n$ steps. 
At each step, the particle moves a distance $b$ either to the right or to the left, with each step taking a time interval $\tau$, so that the jump rate is given by $\gamma = 1/\tau$. 
The evolution of $W(x, x_{\rm i}, n)$ is governed by the recurrence relation,~\cite{Kac_47}
\begin{align}
W(x, x_{\rm i}, n+1) = \frac{1}{2} W(x-b, x_{\rm i}, n) + \frac{1}{2} W(x+b, x_{\rm i}, n) .
\label{eq:unbiased_A}
\end{align}
We denote the solution $W(x, x_{\rm i}, n)$ without any boundary conditions as $W_0(x, x_{\rm i}, n)$. 
Subtracting $W_0(x, x_{\rm i}, n)$ from both sides of Eq.~(\ref{eq:unbiased_A}) and dividing by $\tau$, 
we obtain an equivalent expression in terms of the transition probability $G_0(x, x_{\rm i}, n\tau)$ with $t=n\tau$,
\begin{align}
\frac{G_0(x, x_{\rm i}, t+\tau) - G_0(x, x_{\rm i}, t)}{\tau} =
\frac{\gamma}{2} G_0(x-b, x_{\rm i}, t) - \gamma G_0(x, x_{\rm i}, t) + \frac{\gamma}{2} G_0(x+b, x_{\rm i}, t) .
\label{eq:unbiased_A1}
\end{align}
If the limit $\tau \rightarrow 0$ is taken on the left-hand side of Eq.~(\ref{eq:unbiased_A1}) while keeping $\gamma$ finite, the expression reduces to Eq.~(\ref{eq:unbiased}). ~\cite{Montroll_67,Redner_01}
Strictly speaking, however, taking the limit $\tau \rightarrow 0$ implies that $\gamma = 1/\tau$ diverges; 
discrete-time random walks (DTRWs) and continuous-time random walks (CTRWs) should be distinguished unless a steady state is reached, in which the dependence on $n$ or $t$ disappears.  \cite{RAZINAQVI_82}
Therefore, we study the combined generating function 
and Fourier transform (lattice Green function) corresponding to Eq.~(\ref{eq:unbiased_A}). ~\cite{Montroll_67,Hughes_95,Weiss_94,Redner_01}

We first define the generating function as $\hat{W}_0(x, x_{\rm i}, z) = \sum_{n \geq 0} z^n W_0(x, x_{\rm i}, n)$.
Applying $\sum_{n \geq 0} z^n$ to both sides of Eq.~(\ref{eq:unbiased_A}), and using the identity
\[
\sum_{n \geq 0} z^n W_0(x, x_{\rm i}, n+1) = \sum_{n \geq 0} z^{n-1} W_0(x, x_{\rm i}, n) - \frac{1}{z} W_0(x, x_{\rm i}, 0),
\]
we obtain
\begin{align}
\hat{W}_0(x, x_{\rm i}, z) - W_0(x, x_{\rm i}, 0) = z \left( \frac{1}{2} \hat{W}_0(x - b, x_{\rm i}, z) + \frac{1}{2} \hat{W}_0(x + b, x_{\rm i}, z) \right).
\label{eq:unbiased_A2}
\end{align}
We next introduce the discrete Fourier transform, ~\cite{Hughes_95}
\begin{align}
\tilde{W}_0(k, x_{\rm i}, z) = \sum_x e^{ikx} \hat{W}_0(x, x_{\rm i}, z),
\label{eq:unbiased_A3}
\end{align}
with its inverse given by
\begin{align}
\hat{W}_0(x, x_{\rm i}, z) = \frac{b}{2\pi} \int_{-\pi/b}^{\pi/b} dk\, e^{-ikx} \tilde{W}_0(k, x_{\rm i}, z).
\label{eq:unbiased_A4}
\end{align}
Applying $\sum_x e^{ikx}$ to both sides of Eq.~(\ref{eq:unbiased_A2}), we obtain
\begin{align}
\tilde{W}_0(k, x_{\rm i}, z) - e^{ikx_{\rm i}} = z \cos(kb) \tilde{W}_0(k, x_{\rm i}, z),
\label{eq:unbiased_A5}
\end{align}
where we have used the initial condition $W_0(x, x_{\rm i}, 0) = \delta_{x, x_{\rm i}}$.
Rearranging Eq.~(\ref{eq:unbiased_A5}) and substituting into Eq.~(\ref{eq:unbiased_A4}), we obtain ~\cite{Montroll_67,Hughes_95,Weiss_94,Redner_01}
\begin{align}
\hat{W}_0(x, x_{\rm i}, z) = \frac{b}{2\pi} \int_{-\pi/b}^{\pi/b} dk\, \frac{\cos[k(x - x_{\rm i})]}{1 - z\cos(kb)}.
\label{eq:unbiased_A6}
\end{align}
The time-domain expression $W_0(x, x_{\rm i}, n)$ can be recovered using either of the following methods,~\cite{Weiss_94}
\begin{align}
W_0(x, x_{\rm i}, n) = \left. \frac{1}{n!}\frac{d^n \hat{W}_0(x, x_{\rm i}, z)}{dz^n} \right|_{z = 0} 
\quad \text{or} \quad
W_0(x, x_{\rm i}, n) =\frac{1}{2\pi i} \oint \frac{\hat{W}_0(x, x_{\rm i}, z)}{z^{n+1}}\,dz,
\label{eq:unbiased_A7}
\end{align}
where the contour of integration is a closed loop enclosing the origin.

Equation~(\ref{eq:unbiased_A6}) has the same form as Eq.~(\ref{eq:G0_0dAg}); therefore, results derived using Eq.~(\ref{eq:G0_0dAg}), regardless of the specific form of $\hat{\varphi}(s)$ or $\hat{\psi}(s)$, should also hold for those obtained from Eq.~(\ref{eq:unbiased_A6}).

Here, we show that
\begin{align}
W_{\rm rfl}(x, x_{\rm i}, t) \equiv W_0(x, x_{\rm i}, t) + W_0(x, -x_{\rm i} - 2m, t),
\label{eq:unbiased_A8}
\end{align}
with $m = 1/2$, is a solution of Eq.~(\ref{eq:unbiased_A}) under the reflecting boundary condition given by~\cite{Feller_70}
\begin{align}
W(0, x_{\rm i}, n+1) = \frac{1}{2} W(b, x_{\rm i}, n) + \frac{1}{2} W(0, x_{\rm i}, n).
\label{eq:unbiased_A9}
\end{align}
The initial condition is given by $\delta_{x, x_{\rm i}}$ with $x_{\rm i} \geq 0$, and we consider a random walk over the domain $x = 0, b, 2b, \ldots$.  
Alternatively, if the walk is extended to $x = -b, 0, b, 2b, \ldots$, the reflecting boundary condition can be equivalently expressed as an auxiliary condition at site $-b$,~\cite{Feller_70}
\begin{align}
W(-b, x_{\rm i}, n) = W(0, x_{\rm i}, n).
\label{eq:unbiased_A10}
\end{align}
By adding Eq.~(\ref{eq:unbiased_A10}) to Eq.~(\ref{eq:unbiased_A}) evaluated at $x = 0$, we recover the reflecting boundary condition in Eq.~(\ref{eq:unbiased_A9}).

Using the generating function $\hat{W}_0(x, x_{\rm i}, z)$ defined in Eq.~(\ref{eq:unbiased_A6}),
we verify that the condition $\hat{W}_{\rm rfl}(-b, x_{\rm i}, z) = \hat{W}_{\rm rfl}(0, x_{\rm i}, z)$ is satisfied when $m = 1/2$. 
This follows from the same reasoning used to show that Eq.~(\ref{eq:unbiasedrfllmb}) is equivalent to Eq.~(\ref{eq:unbiasedrfll0}) when $m = b/2$.
This result is consistent with the solution given on p.~369 of Ref.~\citenum{Feller_70}, by considering that the random walk has possible positions $1,2,3, \cdots $. 
The shift $m = b/2$ also matches that used in Eq.~(\ref{eq:Grflsol}), which was obtained within the CTRW framework.

\renewcommand{\theequation}{B.\arabic{equation}}
\setcounter{equation}{0}  
\section*{Appendix B. Correction to Chandrasekhar’s Expression}

Chandrasekhar's expression, given by 
\begin{align}
G_{\rm C}(x,x_{\rm i},t)\equiv G_0(x,x_{\rm i}, t)+G_0(x,-x_{\rm i}, t) ,
\label{eq:Chand0}
\end{align}
does not satisfy the reflecting boundary condition [Eq. (\ref{eq:ubnaturalbc}) or equivalently Eq. (\ref{eq:ubaux})]. 
However, Chandrasekhar's expression can be corrected to satisfy the reflecting boundary condition by considering the correction terms in the Laplace domain as 
\cite{Montroll_67,Schwarz_75,SekiJCP_23,Seki_23}, 
\begin{align}
\hat{G}_{\rm Cc}(x,x_{\rm i},s)= \hat{G}_{\rm C}(x,x_{\rm i},s) +\sum_{j=0}^\infty \hat{G}_0(x,-x_{\rm i}-bj, s) \hat{\zeta}(j,s),
\label{eq:Chand}
\end{align}
where $\hat{\zeta}(j,s)$ will be determined below. 
Using $G_0(x-x_{\rm i}, t)\equiv G_0(x,x_{\rm i}, t)$ [Eq. (\ref{eq:G0_0t_v1})], 
we express, 
\begin{align}
\sum_{j=0}^\infty \hat{G}_0(x_{\rm i}+bj, s) \hat{\zeta}(j,s)=\sum_{j=0}^\infty \hat{G}_0(-b+x_{\rm i}+bj, s) \hat{\zeta}(j-1,s) -
\hat{G}_0(-b+x_{\rm i}, s) \hat{\zeta}(-1,s) .
\label{eq:zeta}
\end{align}
By considering the reflecting boundary 
[$G_{\rm Cc}(-b,x_{\rm i},t)=G_{\rm Cc}(0,x_{\rm i},t)$] and 
Eq. (\ref{eq:zeta}), we note that 
$\hat{\zeta}(j,s)=\hat{\zeta}(j-1,s)=\hat{\zeta}(-1,s)$ and 
\begin{align}
\hat{\zeta}(-1,s)=\hat{G}_{\rm n}(x_{\rm i}, s)/\hat{G}_0(-b+x_{\rm i}, s),
\label{eq;zetam1}
\end{align}
where 
\begin{align}
\hat{G}_{\rm n}(x_{\rm i}, s)=2\hat{G}_0(x_{\rm i}, s)-\hat{G}_0(x_{\rm i}+b, s)-\hat{G}_0(x_{\rm i}-b, s), 
\label{eq;Gn}
\end{align}
should be satisfied. 
$\hat{G}_{\rm n}(x_{\rm i}, s)$ can be expressed as
\begin{align}
\hat{G}_{\rm n}(x_{\rm i}, s)= \frac{b}{\pi} \int_{-\pi/b}^{\pi/b} dk\, \frac{\cos(k x_{\rm i})\left[1-\cos\left(kb \right)\right]}{s+\gamma\left[1-\cos\left(kb \right)\right]} ,
\label{eq:Gn1}
\end{align}
using $\cos(A)+\cos(B)=2 \cos[(A+B)/2]\cos[(A-B)/2]$. 
Therefore, we obtain Chandrasekhar's expression corrected to satisfy the reflecting boundary condition as, 
\begin{align}
\hat{G}_{\rm Cc}(x,x_{\rm i},s)= \hat{G}_{\rm C}(x,x_{\rm i},s) +\sum_{j=0}^\infty \hat{G}_0(x,-x_{\rm i}-bj, s) \hat{G}_{\rm n}(x_{\rm i}, s)/\hat{G}_0(-b+x_{\rm i}, s), 
\label{eq:Chandf}
\end{align}
where $\hat{G}_{\rm n}(x_{\rm i}, s)$ is given by Eq. (\ref{eq:Gn1}). 
Both Eqs. (\ref{eq:Grflsol}) and (\ref{eq:Chandf})
represents the transition probability under the reflecting boundary condition at site $0$. 
Equation (\ref{eq:Grflsol}) is invariant but Eq. (\ref{eq:Chandf}) is not invariant under the exchange of $x$ and $x_{\rm i}$. 
Therefore, these are different results. 
By considering that free diffusion is studied, 
the fluctuation theorem is satisfied for Eq. (\ref{eq:Grflsol}) but is broken for Eq. (\ref{eq:Chandf}). \cite{Seki_23}
In the limit of $b \rightarrow 0$, we find $\hat{G}_{\rm n}(x_{\rm i}, s) \rightarrow 0$ and 
$G_{\rm rfl}(0,x_{\rm i}, t) \rightarrow G_{\rm C}(x,x_{\rm i},t)$. 
As a results, they coincide and satisfy the fluctuation theorem in the limit of $b \rightarrow 0$. 

\renewcommand{\theequation}{C.\arabic{equation}}
\setcounter{equation}{0}  
\section*{Appendix C. Correction to Chandrasekhar’s Expression in DTRW}

Chandrasekhar's expression, given by
\begin{align}
W_{\rm C}(x,x_{\rm i},t) \equiv W_0(x,x_{\rm i}, t) + W_0(x,-x_{\rm i}, t),
\label{eq:AppendixC1}
\end{align}
does not satisfy the reflecting boundary condition, $W_{\rm C}(-b, x_{\rm i}, t) \neq W_{\rm C}(0, x_{\rm i}, t)$. 

As in Appendix B for the CTRW case, Chandrasekhar's expression can be modified to satisfy the reflecting boundary condition by introducing correction terms,
\begin{align}
\hat{W}_{\rm Cc}(x,x_{\rm i},z) = \hat{W}_{\rm C}(x,x_{\rm i},z) + \sum_{j=0}^\infty \hat{W}_0(x,-x_{\rm i} - bj, z)\, \hat{\zeta}(j,z),
\label{eq:AppendixC2}
\end{align}
where the coefficients $\hat{\zeta}(j,z)$ will be determined below.
Imposing the reflecting boundary condition $W_{\rm Cc}(-b, x_{\rm i}, t) = W_{\rm Cc}(0, x_{\rm i}, t)$, we find that $\hat{\zeta}(j,z) = \hat{\zeta}(j-1,z) = \hat{\zeta}(-1,z)$ and
\begin{align}
\hat{\zeta}(-1,z) = \frac{\hat{W}_{\rm n}(x_{\rm i}, z)}{\hat{W}_0(-b + x_{\rm i}, z)},
\label{eq:AppendixC3}
\end{align}
where $\hat{W}_{\rm n}(x_{\rm i}, z)$ is given by
\begin{align}
\hat{W}_{\rm n}(x_{\rm i}, z) = \frac{b}{\pi} \int_{-\pi/b}^{\pi/b} dk\, \frac{\cos(k x_{\rm i}) \left[1 - \cos(kb)\right]}{1 - z \cos(kb)}.
\label{eq:AppendixC4}
\end{align}
Substituting this into Eq.~(\ref{eq:AppendixC2}), we obtain the corrected Chandrasekhar expression that satisfies the reflecting boundary condition:
\begin{align}
\hat{W}_{\rm Cc}(x,x_{\rm i},z) = \hat{W}_{\rm C}(x,x_{\rm i},z) + \sum_{j=0}^\infty \hat{W}_0(x,-x_{\rm i} - bj, z)\, \frac{\hat{W}_{\rm n}(x_{\rm i}, z)}{\hat{W}_0(-b + x_{\rm i}, z)}.
\label{eq:AppendixC5}
\end{align}

Both Eq.~(\ref{eq:unbiased_A8}) with $m = 1/2$ and Eq.~(\ref{eq:AppendixC5}) represent transition probabilities under the reflecting boundary condition at site $x = 0$. However, Eq.~(\ref{eq:unbiased_A8}) with $m = 1/2$ is symmetric under the exchange of $x$ and $x_{\rm i}$, while Eq.~(\ref{eq:AppendixC5}) is not. Therefore, they represent different solutions.
Since the problem involves free diffusion, the fluctuation theorem holds for Eq.~(\ref{eq:unbiased_A8}) with $m = 1/2$, but it is violated for Eq.~(\ref{eq:AppendixC5}).

\renewcommand{\theequation}{D.\arabic{equation}}  
\setcounter{equation}{0}  
\section*{Appendix D. Reflecting boundary condition in asymmetric (biased) random walk}
We study one-dimensional asymmetric random walk,  
\begin{align}
\frac{\partial}{\partial t} G(x,x_{\rm i}, t)=\Gamma_{\rm p} G(x-b,x_{\rm i}, t)-
\left(\Gamma_{\rm p}+\Gamma_{\rm m}  \right) G(x,x_{\rm i}, t)+
\Gamma_{\rm m}  G(x+b,x_{\rm i}, t) ,
\label{eq:biased0}
\end{align}
where $G(x,x_{\rm i}, t)$, $\Gamma_{\rm p}$, and $\Gamma_{\rm m}$ indicate the transition probability function, the transition rate constant to the positive direction 
and that in the negative direction of the $x$-coordinate; 
$\Gamma_{\rm p}=\Gamma_{\rm m}=\gamma/2$ for the symmetric random walk. 
The reflecting boundary condition at site $0$ can be expressed as:  \cite{vanKampen}
\begin{align}
\frac{\partial}{\partial t} G_{\rm rfl}(0,x_{\rm i}, t)=-\Gamma_{\rm p} G_{\rm rfl}(0,x_{\rm i}, t)+\Gamma_{\rm m} G_{\rm rfl}(b,x_{\rm i}, t) .
\label{eq:reflboundaryc}
\end{align} 
The initial condition is given by $\delta_{x, x_{\rm i}}$ with $x_{\rm i}\geq 0$ and we consider a random walk over $x=,0,b,2b,\cdots$. 
By considering a random walk over $x=-b,0,b,2b,\cdots$, 
the reflecting boundary condition at size $0$ can be taken into account by considering  
an auxiliary boundary condition at site $-b$, \cite{vanKampen}
\begin{align}
\Gamma_{\rm p} G(-b,x_{\rm i}, t)=
\Gamma_{\rm m} G(0,x_{\rm i}, t). 
\label{eq:baux}
\end{align}
We will obtain exact solutions of Eq. (\ref{eq:biased0}) with the initial condition, 
$\delta_{x, x_{\rm i}}$ for $x_{\rm i}\geq 0$, under the reflecting boundary condition [Eq. (\ref{eq:reflboundaryc})] 
by considering a random walk over $x=-b,0,b,2b,\cdots$, and imposing the boundary condition at site $-b$ using Eq. (\ref{eq:baux}).

It is convenient to introduce 
$g(x,x_{\rm i},t)$, which is not altered by exchanging $x$ and $x_{\rm i}$  \cite{Seki_23}
\begin{align}
G (x,x_{\rm i},t)&=\left(\Gamma_{\rm p} /\Gamma_{\rm m}  \right)^{(x-x_{\rm i})/(2b)}
 g(x,x_{\rm i},t) .
\label{eq:refl0_st}
\end{align}
Using Eq. (\ref{eq:biased0}), we obtain the time-evolution equation for $g(x,x_{\rm i},t)$ as, \cite{Seki_23}
\begin{align}
\frac{\partial}{\partial t} g(x,x_{\rm i}, t)=\sqrt{\Gamma_{\rm p} \Gamma_{\rm m}}g(x-b,x_{\rm i}, t)-
\left(\Gamma_{\rm p}+\Gamma_{\rm m}  \right) g(x,x_{\rm i}, t)+
\sqrt{\Gamma_{\rm p} \Gamma_{\rm m}} g(x+b,x_{\rm i}, t) ,
\label{eq:g01}
\end{align}
which is unaltered by exchanging $\Gamma_{\rm p}$ and $\Gamma_{\rm m}$  
and therefore remains invariant when the direction of x-axis is changed; 
$g(x,x_{\rm i},t)$ is not altered by exchanging $x$ and $x_{\rm i}$. 
The solution without boundary condition 
 can be expressed in the Laplace domain using 
 $\hat{g}_0(|x-x_{\rm i}|,s)=\hat{g}(x,x_{\rm i},s)$ as  
\begin{align}
\hat{g}_0(|x-x_{\rm i}|,s)&= \frac{b}{2\pi} \int_{-\pi/b}^{\pi/b} dk\, \frac{\exp[-ik(x-x_{\rm i})]}{s+\Gamma_{\rm p}+\Gamma_{\rm m} 
-2\sqrt{\Gamma_{\rm p} \Gamma_{\rm m}}\, \cos(k b)}.
\label{eq:abs_8_1}
\end{align}
$\hat{g}_0(x-x_{\rm i},s)$ is an even function of $x-x_{\rm i}$. 
The auxiliary boundary condition is given by 
\begin{align}
\Gamma_{\rm p}^{1/2}    g(-b,x_{\rm i},t) =\Gamma_{\rm m}^{1/2}  g(0,x_{\rm i},t) .
\label{eq:auxg}
\end{align}
As in Appendix B, we introduce an expression, 
\begin{align}
\hat{g} (x,x_{\rm i},s)=\hat{g}_{\rm rfl}(x,x_{\rm i},s)+
\sum_{j=0}^\infty  \hat{g}_0(x+x_{\rm i}+b j+\ell_2,s)\hat{\xi}(j,s),  
\label{eq:g1}
\end{align}
where we introduced 
\begin{align}
\hat{g}_{\rm rfl}(x,x_{\rm i},s)=\hat{g}_0(|x-x_{\rm i}|,s)+
\hat{g}_0(|x+x_{\rm i}+\ell_1 |,s)
\label{eq:grefl}
\end{align}
and $\hat{\xi} (j,s)$ is determined using the boundary condition given by Eq. (\ref{eq:auxg}); 
$\ell_1$ and $\ell_2$ are constants. 

By using, 
\begin{multline}
\sum_{m=0}^\infty g_0(x_{\rm i} +bm+\ell_2,t)\hat{\xi}(m,s)
= \sum_{j=0}^\infty g_0(-b+x_{\rm i} +bj+\ell_2,t)\hat{\xi}(j-1,s) -\\
g_0(-b+x_{\rm i}+\ell_2 ,t)\hat{\xi}(-1,s) ,
\label{eq:sumubl}
\end{multline}
where $j=m+1$ is introduced, 
and Eq. (\ref{eq:auxg}), we find, 
\begin{align}
\hat{\xi}(j,s)/\hat{\xi}(j-1,s)=\sqrt{\Gamma_{\rm m}/\Gamma_{\rm p}}  .
\label{eq:xi}
\end{align}
We also note that the following relation should hold, 
\begin{align}
\Gamma_{\rm p}^{1/2}  \hat{g}_{\rm rfl}(-b,x_{\rm i},s)
=
\Gamma_{\rm m}^{1/2}  
\left[\hat{g}_{\rm rfl}(0,x_{\rm i},s)
-\hat{g}_0(x_{\rm i} +\ell_2-b,s)\hat{\xi}(-1,s) 
\right].
\label{eq:15_2_st}
\end{align}
We rearrange (\ref{eq:15_2_st}) and find, 
\begin{align}
\hat{\xi} (-1,s)= \frac{\hat{g}_{\rm rfl}(0,x_{\rm i},s)-\left( \Gamma_{\rm p}/\Gamma_{\rm m}
\right)^{1/2}\hat{g}_{\rm rfl}(-b,x_{\rm i},s)}{g_0(x_{\rm i} +\ell_2-b,s)} . 
\label{eq:15_3_st}
\end{align}
Using $\cos(A)+\cos(B)=2 \cos[(A+B)/2]\cos[(A-B)/2]$, we obtain, 
\begin{align}
\hat{g}_{\rm rfl}(-b,x_{\rm i},s)&= \frac{b}{2\pi} \int_{-\pi/b}^{\pi/b} dk\, \frac{2\cos[k(x_{\rm i} + \ell_1/2)]\cos[k(b- \ell_1/2)]}{s+\Gamma_{\rm p}+\Gamma_{\rm m} 
-2\sqrt{\Gamma_{\rm p} \Gamma_{\rm m}}\, \cos(k b)}, 
\label{eq:biasedrfllmb}\\
\hat{g}_{\rm rfl}(0,x_{\rm i},s)&= \frac{b}{2\pi} \int_{-\pi/b}^{\pi/b} dk\, \frac{2\cos[k(x_{\rm i} + \ell_1/2)]\cos(k \ell_1/2)}{s+\Gamma_{\rm p}+\Gamma_{\rm m} 
-2\sqrt{\Gamma_{\rm p} \Gamma_{\rm m}}\, \cos(k b)}.  
\label{eq:biasedrfll0}
\end{align}
By setting $\ell_1=b$, we have $\hat{g}_{\rm rfl}(-b,x_{\rm i},s)=\hat{g}_{\rm rfl}(0,x_{\rm i},s)$.

Using Eq. (\ref{eq:xi}) and $\ell_1=b$, we find $\hat{\xi}(j,s)=(\Gamma_{\rm m} /\Gamma_{\rm p})^{(j+1)/2} \hat{\xi}(-1,s)$ 
and obtain, 
\begin{align}
\hat{\xi}(j,s)=  \frac{\hat{g}_{\rm rfl}(0,x_{\rm i},s)}{\hat{g}_0(x_{\rm i} +\ell_2-b,s)}
\left( \frac{\Gamma_{\rm m}}{\Gamma_{\rm p}}
\right)^{(j+1)/2}
 \left[1
-\left( \frac{\Gamma_{\rm p}}{\Gamma_{\rm m} }
\right)^{1/2}
\right] , 
\label{eq:xi_j}
\end{align}
where $\hat{g}_0(x_{\rm i} +\ell_2-b,s)$ and $\hat{g}_{\rm rfl}(0,x_{\rm i},s)$ can be expressed as 
\begin{align}
\hat{g}_0(x_{\rm i} +\ell_2-b,s)&= \frac{b}{2\pi} \int_{-\pi/b}^{\pi/b} dk\, \frac{\cos[k\left(x_{\rm i} +\ell_2-b\right)]}{s+\Gamma_{\rm p}+\Gamma_{\rm m} 
-2\sqrt{\Gamma_{\rm p} \Gamma_{\rm m}}\, \cos(k b)} ,
\label{eq:g0mi}\\
\hat{g}_{\rm rfl}(0,x_{\rm i},s)&= \frac{b}{2\pi} \int_{-\pi/b}^{\pi/b} dk\, \frac{2\cos[k(x_{\rm i} + b/2)]\cos(kb/2)}{s+\Gamma_{\rm p}+\Gamma_{\rm m} 
-2\sqrt{\Gamma_{\rm p} \Gamma_{\rm m}}\, \cos(k b)}.  
\label{eq:g0m}
\end{align}

By comparing $\cos[k\left(x_{\rm i} +\ell_2-b\right)]$ in Eq. (\ref{eq:g0mi}) and 
$\cos[k(x_{\rm i} + b/2)]$ in Eq. (\ref{eq:g0m}), 
$x_{\rm i}$ dependence can be eliminated in Eq. (\ref{eq:xi_j}) when 
$\ell_2=3b/2$. 
We study Eq. (\ref{eq:refl0_st}) with Eq. (\ref{eq:g1}) by setting $\ell_1=b$ and $\ell_2=3b/2$, 
where we have $\hat{g}_{\rm rfl}(x,x_{\rm i},s)=\hat{g}_0(|x-x_{\rm i}|,t)+g_0(x+x_{\rm i}+b,s)$ and 
\begin{align}
\hat{\zeta}(j,s)= 2\cos(kb/2)\left( \frac{\Gamma_{\rm m}}{\Gamma_{\rm p}}
\right)^{(j+1)/2} 
 \left[1
-\left( \frac{\Gamma_{\rm p}}{\Gamma_{\rm m} }
\right)^{1/2}
\right] . 
\label{eq:15_4_st_1}
\end{align}
Using Eq. (\ref{eq:g1}) with $\ell_1=b$ and $\ell_2=3b/2$, we obtain,
\begin{align}
g (x,x_{\rm i},t)=g_{\rm rfl}(x,x_{\rm i},t)+
\sum_{j=0}^\infty 2\cos(kb/2) g_0(x+x_{\rm i}+b j+3b/2,t)\left( \frac{\Gamma_{\rm m}}{\Gamma_{\rm p}}
\right)^{(j+1)/2} 
 \left[1
-\left( \frac{\Gamma_{\rm p}}{\Gamma_{\rm m} }
\right)^{1/2}
\right] ,  
\label{eq:gf}
\end{align}
where $g_{\rm rfl} (x,x_{\rm i},t)$ in the Laplace space is given by Eq. (\ref{eq:grefl}). 
$G (x,x_{\rm i},t)$ can be obtained using Eqs. (\ref{eq:refl0_st}) and (\ref{eq:gf}). 

An alternative solution can be found by setting, 
\begin{align}
\hat{g}_{\rm rfla}(x,x_{\rm i},s)=\hat{g}_0(|x-x_{\rm i}|,s)+\sqrt{\Gamma_{\rm p}/\Gamma_m}\, 
\hat{g}_0(|x+x_{\rm i}+\ell_1 |,s) , 
\label{eq:grefla}
\end{align}
where $\ell_1=b$, instead of Eq. (\ref{eq:grefl}). 
Then, we find Eq. (\ref{eq:15_3_st}) and, 
\begin{align}
\hat{g}_{\rm rfla}(0,x_{\rm i},s)-\left( \Gamma_{\rm p}/\Gamma_{\rm m}
\right)^{1/2}\hat{g}_{\rm rfla}(-b,x_{\rm i},s) =\left(1-\Gamma_{\rm P}/\Gamma_{\rm m}\right)\hat{g}_0(x_{\rm i},s) ,
\label{eq:reflanu}
\end{align}
by setting $\ell_1=b$ in the numerator of the right-hand side of Eq. (\ref{eq:15_3_st}), 
where we have $\hat{g}_{\rm rfla}(x,x_{\rm i},s)$ instead of $\hat{g}_{\rm rfl}(x,x_{\rm i},s)$. 
Using $\ell_2=b$ and 
$\hat{\xi}(j,s)=(\Gamma_{\rm m} /\Gamma_{\rm p})^{(j+1)/2} \hat{\xi}(-1,s)$, we obtain \cite{Montroll_67,KOSZTOLOWICZ_98}, 
\begin{align}
g (x,x_{\rm i},t)=g_{\rm rfla}(x,x_{\rm i},t)+
\frac{\Gamma_{\rm m}-\Gamma_{\rm p}}{\sqrt{\Gamma_{\rm p}\Gamma_{\rm m}}}
\sum_{j=0}^\infty g_0(x+x_{\rm i}+b j+b,t)\left( \frac{\Gamma_{\rm m}}{\Gamma_{\rm p}}
\right)^{j/2} .   
\label{eq:gfa}
\end{align}

When a particle with the charge $q$ executes random walk under bias electric field denoted by $F$, 
the energy difference between the adjacent sites can be given by
$q F b$ and the detailed balance condition can be expressed as
$\Gamma_{\rm p}/\Gamma_{\rm m} =\exp\left[qFb/\left(
k_{\rm B} T
\right)
\right]$, 
where $T$ and $k_{\rm B}$ indicate the temperature and the Boltzmann constant. \cite{vanKampen}
Because $\hat{g} (x,x_{\rm i},s)$ is unaltered by exchanging $x$ and $x_{\rm i}$ 
in both Eqs. (\ref{eq:gf}) and (\ref{eq:gfa}), 
we find the fluctuation relation given by, 
\begin{align}
\frac{G(x,x_{\rm i}, t)}{G(x_{\rm i},x, t)}=\left(\frac{\Gamma_{\rm p}}{\Gamma_{\rm n}}
\right)^{(x-x_{\rm i})/b} , 
\label{eq:asymflu}
\end{align}
when the detailed balance condition is satisfied. 
However, the fluctuation relation is broken for the transition probability 
obtained by setting other values of $\ell_1$ and $\ell_2$ including the case of  $\ell_1=\ell_2=0$ 
in Eqs. (\ref{eq:g1}) and (\ref{eq:grefl}). \cite{Seki_23}
When the continuous limit in space is taken, a unique solution 
satisfying the fluctuation relation is obtained. \cite{Seki_23,SekiJCP_23}

Finally, we show that the explicit expression of $g_0(|x-x_{\rm i}|,t)$ can be obtained using Eq. (\ref{eq:abs_8_1}) as 
\begin{align}
g_0(|x-x_{\rm i}|,t)&= \frac{b}{2\pi} \int_{-\pi/b}^{\pi/b} dk\, \exp[-ik(x-x_{\rm i})]
\exp\left[- \left(\Gamma_{\rm p}+\Gamma_{\rm m} 
-2\sqrt{\Gamma_{\rm p} \Gamma_{\rm m}}\, \cos(k b) 
\right)t
\right] \\
&= \exp\left[- \left(\Gamma_{\rm p}+\Gamma_{\rm m} \right)t\right]  I_{|x-x_{\rm i}|/b}\left(2\sqrt{\Gamma_{\rm p} \Gamma_{\rm m}}\, t \right) , 
\label{eq:g0t}
\end{align}
where the integral representation of the modified Bessel function of the first kind [$I_n (z)$] for an integer index is used. 
Substituting Eq. (\ref{eq:g0t})  into Eq. (\ref{eq:refl0_st}) yields the known expression given in Eq. (154) of Ref. \citenum{Montroll_67}.

\end{document}